\documentclass[12pt,preprint]{aastex}

\shorttitle{HI in WLM}
\shortauthors{Kepley et al.}

\def\ha{H$\alpha \ $}
\def\kms{km~s$^{-1}\ $}
\def\Msun{M$_\odot\ $}

\begin{document}

\title{A High-Resolution Study of the HI Content of \\ Local Group Dwarf Irregular Galaxy WLM}

\author{Amanda A. Kepley and Eric M. Wilcots} 
\affil{Washburn Observatory, University of Wisconsin, 475 North Charter Street, Madison, WI 53706} 
\email{kepley@astro.wisc.edu,ewilcots@astro.wisc.edu}

\author{Deidre A. Hunter}
\affil{Lowell Observatory, 1400 West Mars Hill Road, Flagstaff, AZ 86001}
\email{dah@lowell.edu}

\author{Tyler Nordgren}
\affil{Department of Physics, University of Redlands, 1200 East Colton Avenue, Redlands, CA 92373}
\email{tyler\_nordgren@redlands.edu}

\begin{abstract}

Dwarf irregular galaxies are unique laboratories for studying the
interaction between stars and the interstellar medium in low mass
environments. We present the highest spatial resolution observations
to date of the neutral hydrogen content of the Local Group dwarf
irregular galaxy WLM. We find that WLM's neutral hydrogen distribution
is typical for a galaxy of its type and size and derive an HI mass of
$6.3 \pm 0.3 \times 10^7 \ $\Msun for WLM. In addition, we derive an
HI extent for WLM of $\sim 30\arcmin$, which is much less than the
$45\arcmin$ extent found by \citet{hsm81}. We show that the broken
ring of high column density neutral hydrogen surrounding the center of
WLM is likely the result of star formation propagating out from the
center of the galaxy.  The young stars and \ha emission in this galaxy
are mostly correlated with the high column density neutral
hydrogen. The gap in the central ring is the result of star formation
in that region using up, blowing out, or ionizing all of the neutral
hydrogen. Like many late-type galaxies, WLM's velocity field is
asymmetric with the approaching (northern) side appearing to be warped
and a steeper velocity gradient for the approaching side than for the
receding side in the inner region of the galaxy. We derive a dynamical
mass for WLM of $2.16 \times 10^9\ \rm{M_\odot}$.

\end{abstract}


\section{Introduction}\label{sec:introduction}

The interstellar medium of dwarf irregular galaxies provides important
information about how stars and gas interact in low mass environments.
Feedback from massive stars is the dominant process that affects the
interstellar medium in these galaxies. Their low mass makes it easier
for violent episodes of star formation to disrupt the galaxy's gas,
possibly even expelling it from the galaxy's potential
\citep{maclow1999}. The lack of differential rotation in dwarf
irregulars contributes to longer lifetimes for features like holes in
the interstellar medium created by star formation. These longer hole
lifetimes allow one to probe the interaction of massive stars and the
interstellar medium over longer timescales than possible in other
types of galaxies.

As part of an ongoing survey of dwarf irregular galaxies with
apparently extended neutral hydrogen emission
\citep{2002AJ....123.1476W,hunter02}, we have obtained observations of
the neutral hydrogen, ionized hydrogen, and stellar content of an
isolated Local Group dwarf irregular galaxy called WLM.  Its nearest
neighbor, the Cetus dwarf spheroidal galaxy, is approximately 200 kpc
away, assuming a distance to Cetus dSph of 775 kpc
\citep{1999AJ....118.2767W}. Half of WLM's stars, by mass, were formed
over 9 Gyr ago, and since then it has been forming stars at a
leisurely rate of $1 - 2 \times 10^{-4} \ \rm{M_\odot} \ \rm{yr}^{-1}$
\citep{dolphin00}. In the last Gyr or so, there has been a spurt of
star formation, with new stars being formed at three times the average
rate in the past \citep{dolphin00}.  \citet{youngblood99} identified
30 \ha regions in WLM, which is consistent with its recent spurt of
star formation. Like many dwarf irregulars, WLM has a low metallicity;
its oxygen abundance is $12+\log {\rm O/H}=7.78$--8.23
\citep{skillman89,hunter99}, where 8.7 is solar. Given its
metallicity, it is not surprising that \citet{taylor01} did not detect
any CO emission in this galaxy, since the CO to $\rm{H_2}$ ratio is
thought to decrease significantly for low metallicity galaxies. WLM
also has a globular cluster with an age of $14.8 \pm 0.6 \ \rm{Gyr}$
on its western edge \citep{1999ApJ...521..577H}, which is very unusual
for a faint dwarf irregular \citep{billett03}.

We present the highest spatial resolution study to date of the neutral
hydrogen content of WLM. Previous studies include \citet{jackson04}
(spatial resolution: 39$\arcsec$ by 161$\arcsec$; velocity resolution:
$1.65 \ $\kms), \citet{barnes04} (spatial resolution: 15.5$\arcmin$;
velocity resolution: $1.12 \ $\kms), and \citet{hsm81} (spatial
resolution: 9$\arcmin$; velocity resolution: $11 \ $\kms). WLM's
declination makes it difficult to observe; it is low enough in the sky
for Very Large Array's (VLA) beam to be elongated and is too near the
celestial equator to get good uv coverage with the Australia Telescope
Compact Array (ATCA). We have obtained VLA\footnote{The VLA is part of
the National Radio Astronomy Observatory, which is a facility of the
National Science Foundation operated under cooperative agreement by
Associated Universities, Inc.}  observations of WLM in the B, C, and D
configurations. Although our beam is elongated in the north-south
direction due to the low declination, the high spatial resolution of
these observations allows us to investigate in detail the HI content
of this galaxy. We have complemented these observations with UBV and
\ha imaging to gain a more complete picture of the interaction between
the stars and the interstellar medium in WLM. Our data are discussed
further in Section~\ref{sec:data}. In Section~\ref{sec:himass}, we use
our data to examine the properties of the total HI content of WLM.  We
examine the connection between the small-scale structure of WLM's
interstellar medium and its stars in
Section~\ref{sec:small-scale-hi}. Section~\ref{sec:rotcur} is a
detailed look at the rotation curve of WLM (associated mass modeling
is shown in Appendix~\ref{sec:dynamical_mass}). Our conclusions are
summarized in Section~\ref{sec:summary}.

\section{Data} \label{sec:data}

\subsection{VLA observations} \label{sec:vladata}

Previous observations of WLM \citep{hsm81} indicated that the HI
extent of WLM is larger than the primary beam of the VLA at 20 cm
(45$\arcmin$ versus 30$\arcmin$), so we mapped WLM in D array as a 3
by 3 mosaic with pointing centers $15\arcmin$ apart to make sure that
we would detect the total extent of the large scale structure. The
resulting square field of view is 60$\arcmin$ (at the half-power
points). These observations were taken on 2001 October 19. The data
were on-line hanning smoothed to give a final channel width of 2.57
\kms. The flux calibrator (0137+331) was observed at the beginning and
end of the observations. During the observations, we cycled through
the mosaic positions, observing each for 100~s per cycle. We observed
the phase calibrator (2321-163) once every two cycles through the
mosaic. The total integration time for each pointing was 28
minutes. The effective total integration time for positions near the
center of the mosaic was between 128 and 140 minutes due to the overlap
between the pointings.

To obtain information on the small scale HI structure of WLM, we
observed it using the VLA in B and C configurations. The B array
observations were taken on 2002 July 9 and 12 and consisted of a
single pointing centered on WLM's position.  The flux calibrator
(0137+331) was observed for 15 minutes at the end of each day and the
phase calibrator (2357-114) was observed for 5 minutes after every 30
minutes of time on source. The total on-source integration time for
WLM in B array was 8.5 hours. The C array observations were taken on
2000 March 31 and consisted of a single pointing centered on WLM's
position.  The flux calibrator (0137+331) was observed for 20 minutes
in the middle of the observations and the phase calibrator (2321-163)
was observed for 5 minutes after every 35 minutes on source. The total
on-source integration time was 5.1 hours. The correlator setup for
both the B and C array observations was identical to the D array
observations; the final channel width was 2.57 \kms.

The observations are summarized in Table~\ref{tab:obs_summary}. Data
from each day and each array configuration were calibrated separately
in AIPS using the standard calibration procedure outlined in the AIPS
Cookbook \citep{aipscookbook}. After calibration, the data were
imported into Miriad \citep{sault95} for further flagging as well as
continuum subtraction, imaging, mosaicking, and cleaning.

From these data, we produced three final data cubes: two from the
D-array mosaic and one from the central pointing of the D array mosaic
plus the B and C array observations (hereafter, the B$+$C$+$D array
cube). Table~\ref{tab:data_summary} gives the final noise and beam
size in all three cubes. We use the D array mosaic observations to
investigate the large scale structure and total extent of WLM. After
making an initial image to determine the line-free channels, the
continuum was subtracted from the initial pointings using the task
{\it uvlin}. Then the nine pointings were put together using the
mosaic option in the {\it invert} task. Two different weightings were
used: robust equal to 5 (natural weighting) and robust equal to -5
(uniform weighting). The resulting data cubes were then cleaned using
the {\it mossdi} task to deconvolve the source distribution from the
antenna sampling pattern. Images were cleaned to the 2$\sigma$ noise
level and the main emission region was selected. Both the MEM and
Clean algorithms yielded similar results for these observations. The
Clean algorithm results were used since the final cube had fewer image
artifacts after deconvolution.  Channel maps for this cube are given
in Figure~\ref{fig:chanmap_wlmdnat} and the total intensity (moment
zero) image is shown in Figure~\ref{fig:wlmd_nat_mom0}.

We used the B$+$C$+$D array cube to investigate the small scale HI
structure of WLM. The observations from the B and C configurations
were imaged separately using the {\it invert} task to make sure there
were no problems with the data from individual configurations. After
flagging, the uv data sets for each array configuration were combined
with {\it uvcat}. The resulting data set was continuum subtracted
({\it uvlin}), imaged ({\it invert}), and cleaned to the 2$\sigma$
noise limit ({\it clean} with Miriad choosing the best cleaning
algorithm and the main emission region selected). The final map was
produced using robust equal to zero because this value yielded the
best beam profile; other values yielded non-Gaussian beam
profiles. Figure~\ref{fig:chanmap_wlmbcd} gives channel maps for this
data cube. Figures~\ref{fig:wlm_comb_mom0}, \ref{fig:wlm_comb_mom1},
and \ref{fig:wlm_comb_mom2} show the total intensity (moment zero)
image, the mean velocity (moment one) image, and the velocity
dispersion (moment two) image for this data cube. 

The moment images were created using the AIPS task MOMNT. We smoothed
the cubes to three times the resolution in the RA, Dec, and velocity
directions. The smoothed cube was blanked below the 3$\sigma$ noise
limit and used as a mask for unsmoothed cube. The channels of the
masked, unsmoothed cube are then combined to create a moment zero,
one, or two image of the data.

We did not use the combination of the D array mosaic with the central
B and C array observations in our analysis because the beam (and the
noise) would vary significantly across the image greatly complicating
analysis. A test image created in Miriad combining the D array mosaic
with the B and C observations does not reveal any new galaxy features
that are not apparent in the two cubes described above.

\subsection{Optical Images} \label{sec:optical}

\ha images of WLM were obtained using the Perkins 1.8 m telescope at
Lowell Observatory in 1992 November. The observations used an
800$\times$800 TI CCD provided to Lowell Observatory by the National
Science Foundation, the Ohio State University Fabry-Perot that was
used as a 3:1 focal reducer, an \ha filter with a FWHM of 32 \AA, and
an off-band filter centered at 6440 \AA\ with a FWHM of 95 \AA.  The
off-band filter was used to subtract stellar continuum from the \ha
filter to leave only the \ha nebular emission.  In order to survey the
entire galaxy, two separate pointings were mosaicked to produce the
final \ha image. The exposure times were 3000 s.  The pixel scale was
0.49$\arcsec$, and the seeing was $\sim$2.5$\arcsec$.  Additional
details can be found in \citet{hunter04}.

UBV images of WLM were obtained using a SITe 2048$\times$2048 CCD on
the Hall 1.1 m telescope at Lowell Observatory in three observing runs
during 1998 October, 1999 October, and 2002 November.  The exposures
were $3\times1200$ s in V, $4\times2400$ s in B, and $6\times2400$ s
in U. The seeing was about 3.0$\arcsec$, and the pixel scale was
1.13$\arcsec$.  The telescope position was offset between each image
in order to allow removal of cosmic rays.  The bias was subtracted
using the overscan strip, and the images were flat-fielded using
observations of twilight sky.  \citet{landolt92} standard stars were
used to calibrate the photometry. Additional details can be found in
\citet{hunter06}.

\section{HI Mass and Extent} \label{sec:himass}

We obtained a total HI flux for WLM of $294 \pm 15$ Jy \kms by
integrating the HI spectrum from the naturally weighted D array mosaic
cube. See Figure~\ref{fig:wlm_hi_spectrum} for the spectrum. This HI
flux corresponds to an HI mass of $6.3 \pm 0.3 \times 10^7$ \Msun at
distance of 0.950 Mpc \citep{dolphin00}. Our flux agrees well with the
Parkes Multibeam observations of \citet{barnes04}, which give a total
HI flux of $292^{+25}_{-15}$ Jy \kms. Therefore, our interferometric
observations have recovered most of the HI flux in
WLM. \citet{jackson04} obtained an integrated flux density for WLM of
only 149 Jy \kms from their Australia Telescope Compact Array (ATCA)
observations, which lack the short spacings needed to recover extended
emission.

To compare our naturally weighted, D-array integrated intensity map
(Figure~\ref{fig:wlmd_nat_mom0}) with the single dish maps of
\citet{barnes04} and \citet{hsm81}, we smoothed the D array cube to
the same spatial resolution as \citet[14.4\arcmin]{barnes04} and
\citet[9\arcmin]{hsm81} and created moment maps from the smoothed
cubes. Our smoothed maps reach similar noise levels to both
\citet{barnes04} and \citet{hsm81}.  We see that our map agrees well
with that of \citet{barnes04}. We do not, however, see the irregular,
low surface brightness features observed by
\citet{hsm81}. \citet{hsm81} note that they were not able to separate
emission from WLM from the nearby Magellanic Stream emission because
of their low velocity resolution (11 \kms); the tail of the Magellanic
Stream is very close to the position of WLM at an LSR velocity of
\mbox{--200 \kms} \citep{putman03}. \citet{hsm81} estimated the mass
of WLM by assuming symmetry in WLM's line profile.  Although
\citet{barnes04}'s spatial resolution is larger ($14.4\arcmin$
vs. $9\arcmin$), their superb velocity resolution (1.12 \kms, a factor
of 10 lower than that of \citeauthor{hsm81}) allows one to easily
separate WLM's emission from that of the Magellanic Stream in most
channels.  We, fortunately, resolve out the Magellanic Stream. Given
the line profiles shown in Figure~\ref{fig:wlm_hi_spectrum} and in
\citet{barnes04}, it is clear that WLM's line profile is not symmetric
and that \citet{hsm81}'s integrated HI mass and maps for this galaxy
are not accurate.

WLM's HI only extends to 1.2 times its Holmberg diameter, not the 3.6
times found by \citet{hsm81}. We find that the HI diameters quoted for
other galaxies in the \citet{hsm81} sample are also too large.  Deep
neutral hydrogen observations of Sextans A with the Green Bank
Telescope \citep{bowsher2003} do not detect the extended gas reported
by \citet{hsm81} confirming the results of VLA D-array observations
reported by \citet{2002AJ....123.1476W}. VLA D-array mosaic
observations of NGC 2366 reduced by Wilcots also do not show signs of
the extended HI reported by \citet{hsm81}.

Figure~\ref{fig:wlm_combined_sd_plot} gives the surface density
profile for WLM. The inclination and position angle of the galaxy were
taken from the naturally weighted D array rotation curve (see
Section~\ref{sec:rotcur}) and the disk scale length ($R_D$=2.0\arcmin)
from \citet{hunter06}. Comparing the surface density profile of WLM to
the surface density profiles of the other dwarf galaxies in
Figure~\ref{fig:wlm_combined_sd_plot}, we see WLM's HI distribution is
typical of a dwarf irregular galaxy. Several other galaxies in this
Figure (NGC 2366, NGC 4449, IC 1613, Sextans A, and DDO 88) have
central HI depressions like WLM's. From
Figure~\ref{fig:wlm_combined_sd_plot}, one can see that the average
surface density of WLM's ring is about 1.3 times that of the
center. The contrast increases to 1.4 times that of the center in the
portion of the ring with the highest surface density. This ratio is
smaller than the ratios between ring surface density and central
surface density seen in galaxies like Sextans A ($\sim 1.8$) and DDO
88 ($\sim 2$) and is more like the ratio seen in DDO 105 ($\sim
1.2$). WLM's peak surface density (7.5 \Msun~pc$^{-2}$) is average for
galaxies in this sample. Its HI extends to 8.5 times the optical disk
scale length; six of the ten galaxies in
Figure~\ref{fig:wlm_combined_sd_plot} have larger HI extents.

\section{Small Scale HI Structure} \label{sec:small-scale-hi}

From Figure~\ref{fig:wlm_comb_mom0}, we can see that the neutral
hydrogen distribution of WLM is smooth with a large central depression
and slightly asymmetric. Comparing our Figures~\ref{fig:wlmd_nat_mom0}
and~\ref{fig:wlm_comb_mom0} to Figure 4 in \citet{jackson04}
illustrates why it is crucial to have good uv plane coverage for a
detailed examination of the HI structure of a galaxy. What we were
able to image as a broken ring of high column density HI surrounding
the center of the galaxy, \citet{jackson04} was only able to image as
two asymmetric high column density peaks. We refer to the broken ring
of high column density HI as the ``hook'': the handle of the hook is
the high column density HI extending northward and the hook itself
extends from the handle south and east around the ring until it ends
at the southernmost bit of high column density gas. Approximately 20\%
of the total mass of WLM is in the hook. From the first moment image
(Figure~\ref{fig:wlm_comb_mom1}), which shows the velocity centroid
distribution, and the position-velocity diagram
(Figure~\ref{fig:pv_wlm_comb}), one can see that the velocity
distribution is asymmetric (see Section~\ref{sec:rotcur} for
details). The second moment image in Figure~\ref{fig:wlm_comb_mom2} is
also asymmetric, with the northern half of the galaxy having a higher
average velocity dispersion than the southern half. Note that the
maximum velocity dispersion is approximately 10 \kms and the average
velocity dispersion for the entire galaxy is 4.5 \kms. We do not see
any sign of the hook expanding; plots of spectra in the region as a
function of position are single-peaked.

How does WLM's neutral hydrogen distribution compare to its stellar
and ionized hydrogen distributions? Figure~\ref{fig:wlm_color_contour}
shows a color image of the optical data described in
Section~\ref{sec:optical} with HI contours overlaid. The distribution
of the blue stars is closely correlated with the high column density
HI with the exception of the clump of blue stars on the western edge
of the galaxy. The HI hook is also well correlated with the \ha
emission (see Figure~\ref{fig:wlm_ha_mom0}) except for the C2 complex
in the southwestern portion of the galaxy. This bubble-like \ha region
corresponds to an HI {\em deficit} at the center of the
bubble. Finally, there is faint, diffuse \ha emission along the inner
eastern edge of the hook indicating that there is on-going star
formation in this region.

HI deficits in the center of dwarf irregulars are fairly common (see
Figure~\ref{fig:wlm_combined_sd_plot}).  These rings are thought to be
formed by the stellar winds and supernova explosions of high mass
stars \citep{1987ApJ...317..190M,1994ApJ...427..384E}. The high mass
stars clear out a bubble in the interstellar medium which continues to
expand until its expansion velocity slows to the velocity dispersion
of the galaxy. Gas piles up on the edge of the bubble and can, in
turn, form more stars. There is, however, little definitive
observational evidence that this process is what creates HI
rings. \citet{ott01} find evidence that the central ring in Holmberg~I
was the result of a blowout. They find an age for the HI ring that
matches the age of the stellar populations inside it. Associated
hydrodynamic simulations by \citet{2004A&A...413..939V} show that
multiple supernovae can produce a feature like the ring seen in Ho~I.
HST/ACS observations by \citet{momany05} of the stellar populations of
a portion of SagDIG, however, indicate that there were not enough
stars formed in the central region of the galaxy to generate the
observed HI ring. In their study of DDO 88, \cite{simpson05a} find
that while the star cluster in the center of the galaxy had enough
mass to form the HI ring seen today, the cluster is too old and that
any HI ring formed by a population as old as the cluster would have
dissipated by the present day. Is the hole in the center of WLM the
result of multiple supernova explosions? We can use our observations
of WLM's neutral hydrogen along with the star formation properties of
WLM \citep{dolphin00} to test this hypothesis.

The first quantity we need to estimate is the timescale for the
formation of the hook. Assuming that the hook resulted from an
expanding shell, we can place a rough upper limit on its age. The
average central velocity dispersion of WLM is approximately the final
expansion velocity of the ring if the ring is stalled. Using this
velocity (7 \kms) and the radius of the hook ($R_{inner} = 250 \
\rm{pc}$; $R_{outer}=462 \ \rm{pc}$), we obtain an upper limit on the
expansion time of the ring between 35 Myr (using the inner radius of
the ring) and 69 Myr (using the outer radius of the ring). An upper
limit on the dispersion time for the ring can be estimated by finding
the time it would take for the ring to expand to twice its observed
width. Using the same velocity dispersion as above, this time scale is
approximately 59 Myr. Therefore, the upper limit on the age for WLM's
hook (expansion time plus dispersion time) is between 94 and 128 Myr.

We can use models of a supernova explosion in a uniform medium
\citep{1974ApJ...188..501C} to estimate the energy needed to create
WLM's HI hole. To do this, we must first estimate the number density
of HI at mid-plane ($n_0$). Using an inclination of $75^\circ$ for WLM
and a scale height typical of a dwarf irregular galaxy (500 pc), we
obtain a pathlength through the galaxy of 3800~pc. Dividing the
central surface density ($1.96 \times 10^{21} \ \rm{cm}^{-2}$) by this
pathlength gives an estimate of $n_0$ of $0.16 \ \rm{cm}^{-3}$. A
reasonable upper limit on $n_0$ is 1 cm$^{-3}$ given the properties of
HI in other galaxies. Equation (26) of \citet{1974ApJ...188..501C}
gives an estimate of the supernova energy needed to produce a hole of
radius R from an explosion into a medium of uniform density $n_0$. For
WLM's HI hole, the total HI kinetic energy is between $3.1 \times
10^{51} \ \rm{erg}$ and $2.45 \times 10^{52} \ \rm{erg}$, which
corresponds to between 3 and 24 supernovae.

Has WLM formed enough stars during the last 94 to 128 Myr to produce
that many supernovae? Although there have been several stellar
population studies of WLM
\citep{ferraro89,minniti97,rejkuba00,dolphin00}, there are no
spatially resolved star formation histories for the entire galaxy. The
highest resolution stellar population study was done by
\citet{dolphin00}, who used WFPC2 on Hubble Space Telescope (HST) to
get a detailed star formation history for three regions in the western
half of the galaxy. Figure~\ref{fig:wlm_wfpc2} shows the location of
his fields.\footnote{The caption of Figure 1 in \citet{dolphin00} is
incorrect. The WF2 chip is on the upper left, WF3 is on the lower
left, and WF4 is on the lower right.} Unfortunately, his observations
missed the high column density HI regions, both large HII regions in
the galaxy (\citet{hm95}'s C1 and C2 complexes), and the region of
blue stars uncorrelated with the hook. In his relatively quiescent
survey region, he finds that 85\% of WLM's star formation in the past
200 Myr is in the region surveyed that was closest to the center of
the galaxy and that the galaxy has been experiencing a burst of star
formation in the past Gyr. 

We can see if WLM has had enough recent star formation to form its HI
hole. Using a typical star formation rate for WLM ($1.0 \times 10^{-4}
\ \rm{M}_\odot \ \rm{yr}^{-1}$) scaled to the area interior to the ring
($1.6 \times 10^{-4} \ \rm{M}_\odot \ \rm{yr}^{-1}$) and the maximum
lifetime time for the bubble, we derive a total mass of stars formed
between $1.5 \times 10^4 \ \rm{M}_\odot$ and $2.1 \times 10^4 \
\rm{M}_\odot$. Using a Salpeter initial mass function
\citep{salpeter55}, we find that during the lifetime of the bubble
between 138 and 188 stars were formed with masses greater than 7
M$_\odot$, which exceeds the estimated number of supernovae needed to
form the hole. Therefore, it is plausible that the HI hole in WLM was
caused by star formation in the center of the galaxy.

Unlike DDO 88's unbroken ring, WLM's HI ring has a significant gap on
the western edge of the galaxy and a high column density region
extending north along the major axis of the disk. The handle of the
hook is probably due to an increase in column density because we are
looking through the disk of WLM. The origin of the gap in WLM's HI
ring is more of a mystery. There are bright, blue stars seen near the
C2 \ha complex at the tip of the HI hook, which indicates that in the
past there was more neutral hydrogen in this region than is currently
seen.  It is likely that the star formation in the C2 \ha region has
either used up, blown out, or ionized all of the neutral hydrogen.
Examining spectra as a function of position in the C2 region at the
tip of the HI hook, we see double peaked velocity profiles
characteristic of expanding bubbles with the central channel of the
profile being lower than the two neighboring channels (see
Figure~\ref{fig:wlm_plcub_hole}). The separation between the two peaks
is approximately 9 \kms, which means that the expansion velocity of
the bubble is about 4.5 \kms. This velocity is very similar to the
velocity dispersion of the galaxy as a whole, suggesting that the
bubble is near the end of its expansion. The \ha emission in this
region also exhibits signs of expansion \citep[their Figure 5, slit
position WLM E]{tomita98}. In contrast to the C2 region, the current
round of star formation in the HI hook (indicated by the faint \ha
emission along the inner edge of the hook), however, seems to be just
beginning and it has yet to deplete the neutral hydrogen in this
region. The higher velocity dispersion in the northern half of the
galaxy reflects the disruption of the ISM by the star formation taking
place in the northern half of the galaxy. A detailed, spatially
resolved star formation history of the entire galaxy is needed before
we can attempt to conclusively correlate neutral hydrogen features
with star formation episodes.

\citet{minniti97} have suggested, based on a paper by
\citet{wallin96}, that WLM's globular cluster may have triggered the
recent star formation episode in this galaxy a Gyr ago. Although we
believe that this possibility is unlikely, the globular cluster in
WLM, assuming a typical globular cluster mass of $10^6$ \Msun and a
typical WLM halo velocity of about 40 \kms, does have a total kinetic
energy of $1.6 \times 10^{52} \mathrm{erg}$. If a tenth of this energy
was transferred to WLM, this energy input would be similar to the
energy input by one supernova. Without full orbital parameters for the
globular cluster, we cannot exclude the possibility that a
particularly close/violent encounter between WLM and its globular
cluster has triggered the latest round of star formation.

Another scenario is that a close encounter with the nearby Cetus dwarf
spheroidal galaxy may have triggered the star
formation. Unfortunately, in this case, we are again without detailed
orbital parameters. We can constrain the time since interaction,
however, by assuming that WLM and Cetus dSph are on radial
orbits. Then the time since interaction is just the distance between
the two galaxies (200~kpc) divided by the velocity dispersion of the
Local Group \citep[$\sim 60 \ $ \kms;][]{vandenbergh99}, which yields
a time of 3.4~Gyr. This time is same order of magnitude as the start
of the latest burst of star formation in WLM a Gyr ago
\citep{dolphin00}.

Bars can also create a central HI deficit
\citep[e.g.][]{ryder96}. However, we see no evidence for a bar in a
Spitzer IRAC 3.6 $\mu$m image of the central portion of WLM.  While
the approaching side of WLM is clearly warped, we do not see any
s-shaped isovelocity contours in the center of the galaxy that are
indicative of a bar \citep[e.g.][NGC 1433]{ryder96}.

\section{Rotation Curve} \label{sec:rotcur}

The position-velocity plot of the B$+$C$+$D array data cube in
Figure~\ref{fig:pv_wlm_comb} provides a first look at the rotation
curve of WLM. This plot shows the velocity distribution as a function
of position along a slice corresponding to the major axis of WLM's
neutral hydrogen distribution. There are several things to notice
about this graph. First, WLM is clearly rotating with a velocity of
approximately 30 \kms. Second, the rotation curve is not
symmetric. The rotation curve for the approaching side of the galaxy
increases to a velocity of 30 \kms then levels off; the rotation curve
for the receding side of the galaxy rises much more slowly, does not
level off and is slightly distorted at the location of the
ring. Finally, the ring seen in Figure~\ref{fig:wlm_comb_mom0} is
clearly visible in this plot as the two dark spots at $\pm 1.5
\arcmin$.

To derive a rotation curve for the entire galaxy, we fit a tilted ring
model to the velocity field of the B$+$C$+$D array cube. Tilted ring
models divide the galaxy into a series of annuli and determine the
position angle, inclination, and rotation velocity independently for
each ring.  See \citet{begeman87} for more detailed information about
this type of model. The small beam size of the B$+$C$+$D array cube
allows us to finely sample the rotation curve and minimize beam
smearing effects. Initial guesses for the fit parameters were taken
from the literature. We refined these guesses by fitting functions to
the velocity field. To determine the center of the galaxy and an
approximate inclination, we fit a Brandt type rotation curve to WLM's
velocity field using the {\it GAL} task in AIPS. Then we fit a solid
body rotation curve to the velocity field to determine the systematic
velocity and get a good estimate of the position angle. Note that we
could not use just one function to get all five initial parameters
because although the velocity field is better fit by a solid body
rotation curve, a solid body rotation curve cannot determine the
center of a galaxy.  We used the {\it rotcur} task
\citep{begeman87,begeman89} in GIPSY \citep{gipsy2,gipsy1} to fit a
tilted ring model to the velocity field of the galaxy. The width of
each annulus was $15 \arcsec$. The widths of the annuli were chosen to
be slightly less than the major axis of our elliptical beam. We
started the fits one beam width from the center to avoid beam-smearing
effects. We have corrected the errors produced by {\it rotcur} to take
into account that all the pixels in the ring are not independent.

Note that we also fit tilted ring models to the naturally weighted, D
array cube velocity fields using the same method. However, since these
fits are affected by beam smearing in the inner regions and do not
yield new information about the rotation curve of WLM, we do not
discuss them here.

For the B$+$C$+$D array cube, attempting to fit the inclination,
position angle, and rotation velocity with the same parameters for the
entire galaxy led to large residuals with spatial structure. Since the
approaching (north) and receding (south) halves of WLM are not
symmetric, we decided to fit the inclination, position angle, and
rotation velocity for each side of the galaxy separately. The resulting
fits are shown in Figure~\ref{fig:wlm_rotcur_vripa_all}. The blue and
red lines show the fit for approaching and receding sides of the
B$+$C$+$D array cube, respectively.  The weighted average of the
fitted inclination and position angle for each model are given by thin
solid lines running across the lower two plots.

Examining the top panel of Figure~\ref{fig:wlm_rotcur_vripa_all}, we
see that the rotation curve for the approaching half of the B$+$C$+$D
array data levels off at about $2.5\arcmin$ to a value of 30 \kms,
while the rotation curve of the receding half continues to rise.  The
B$+$C$+$D array rotation curves are relatively bumpy because the
B$+$C$+$D array cube is sampling velocities on size scales of
approximately 50~pc, where small scale velocity features from SNe, for
example, dominate. (See Section~\ref{sec:small-scale-hi} for a
discussion of the velocity dispersion distribution of WLM.)

From Figure~\ref{fig:wlm_rotcur_vripa_all}, one can see that the
inclination of the galaxy based on the high resolution fits is more or
less constant as a function of radius.  The weighted average
inclination for each fit is given in Table~\ref{rotcur_results}.
There is an 8$^\circ$ offset in the inclination between the two sides
of the galaxy for the B$+$C$+$D array rotation curve. 

The position angle as a function of radius shows greater variation
than the inclination as a function of radius. The position angles of
the B$+$C$+$D fits change rather abruptly at particular radii. In the
fit to the approaching side of the B$+$C$+$D array cube, the position
angle changes from about 172$^\circ$ to about 178$^\circ$ at
2.5$\arcmin$, decreases to about 170$^\circ$ at 6.5$\arcmin$ and then
begins to increase again at 8$\arcmin$. Note that the position angle
for radii greater than 9$\arcmin$ is very uncertain. The fit to the
receding side of the B$+$C$+$D array cube increases from about
170$^\circ$ to about 180$^\circ$ between a radius of 5$\arcmin$ and a
radius of 8.5$\arcmin$. The average position angle for the approaching
side is 3$^\circ$ higher than the average position angle for the
receding side for the high resolution data cube. 

The asymmetry in WLM's rotation curve is not unusual; half the dwarf
galaxies in the Westerbork HI Survey of Spiral and Irregular Galaxies
(WHISP) are kinematically lopsided \citep{swaters02}. Only 16\% of
these galaxies are clearly interacting. \citet{swaters02} attributes
the rest of the distortions to long-lived remnants of past
interactions. WLM has at least two neighbors that are close enough to
have disrupted its neutral hydrogen recently: its ancient globular
cluster and the Cetus dwarf spheroidal ($\sim 200$ kpc away). See
Section~\ref{sec:small-scale-hi} for a discussion of the interaction
of WLM with these objects. The presence of a globular cluster suggests
that WLM interacted with something a long time ago; many of the dwarfs
that have young super star clusters in them today have nearby HI
clouds or show other signs of an interaction \citep{billett03}.

We can use our derived rotation curves to fit a mass model to the data
and determine the dynamical mass of WLM. The details of this procedure
are in Appendix~\ref{sec:dynamical_mass}. We derive a dynamical mass
of $2.16 \times 10^9 \ \rm{M_\odot}$ for WLM. Our mass models show
that WLM is extremely dark matter dominated when compared to the dwarf
irregular galaxies in \citet{cote2000}. 

\citet{jackson04} derived a rotation curve for WLM from their low
resolution HI data. Their rotation curve \citep[see][Figure
8]{jackson04} is very similar to our curve (see our
Figure~\ref{fig:wlm_rotcur_vripa_all}). It reaches a velocity of about
38 \kms at a radius of 4 kpc, while our low resolution curve reaches a
velocity of about 42 \kms at the same radius. We also derive similar
halo dynamical masses.\footnote{Note that the dynamical mass quoted in
\citet{jackson04} is incorrect. According to D. Jackson (2006, private
communication), the correct dynamical mass is $1.4 \times 10^9$
\Msun.}

\section{Summary and Conclusions} \label{sec:summary}

Dwarf irregular galaxies are unique laboratories for studying the
interaction between stars and the interstellar medium in low mass
environments. In this paper, we have presented the highest spatial
resolution observations to date of the neutral hydrogen content of the
Local Group dwarf irregular galaxy WLM. These observations have
allowed us to study the interstellar medium of WLM in detail. The
major conclusions of this work are:

\begin{itemize}
\item WLM's neutral hydrogen distribution is typical for a galaxy of
its type and size. Our synthesis observations were able to recover all
of the HI flux from this galaxy; we derive an HI mass of $6.3 \pm 0.3
\times 10^7 \ $\Msun for WLM. We derive an HI extent of $\sim 30
\arcmin$, which is much less than the $45\arcmin$ extent derived by
\citet{hsm81}. In the central region of WLM, the HI has a hook-like
distribution, i.e., a central broken ring of HI with a linear feature
extending northwards from the top of the broken ring.
\item We shoe that the hook is likely the result of star formation
propagating out from the center of the galaxy. The young stars and \ha
emission in this galaxy are mostly spatially correlated with the
neutral hydrogen peaks. The gap on the western edge of the hook is
likely the result of star formation in that region using up, blowing
out, or ionizing all of the neutral hydrogen. The diffuse \ha on the
eastern edge of the hook indicates that there is ongoing star
formation in the central region of WLM.
\item WLM's velocity field is asymmetric. The approaching (northern
 half) appears to be warped and the inner approaching region has a
 steeper velocity gradient than the inner receding region. We derive a
 dynamical mass for WLM of $2.16 \times 10^9 \ \rm{M_\odot}$.
\end{itemize}

In short, WLM is a typical dwarf irregular galaxy undergoing a recent
burst of propagating star formation that is disrupting its neutral
hydrogen. A detailed, spatially resolved star formation history of the
entire galaxy would allow us to track the progress of the star
formation across the galaxy and attempt to correlate particular HI
features with specific episodes of star formation.

\acknowledgments

A. A. K. was supported by a National Science Foundation Graduate
Research Fellowship and a Wisconsin Space Grant Consortium Graduate
Fellowship during portions of this work. E. M. W. gratefully
acknowledges the support of NSF grant AST-0098438. Funding for this
work was provided to D. A. H. by the Lowell Research Fund and the NSF
through grant AST 02-04922. The authors would also like to thank
E. Brinks, B. Elmegreen, D. Westpfahl, V. McIntyre, and G. Ostlin for
their work on the proposal for these observations. A. A. K. would like
to thank Marc Verheijen and Linda Sparke for helpful conversations and
Erwin de Blok for providing us with his WLM data cube.

Facilities: \facility{VLA}, \facility{Hall}, \facility{Perkins}

\appendix

\section{Deriving the Dynamical Mass of WLM} \label{sec:dynamical_mass}

We can use our derived rotation curves to fit a mass model to the data
and determine the dynamical mass of WLM. There are a variety of
techniques and subtleties associated with fitting mass models (see
\citet{swaters2000,deblok2001a,deblok2001b,vandenbosch2001,deblok02,deblok03,mcgaugh03,swaters03a,swaters03b,weldrake03,deblok05,gentile05}
for a sampling). Here we outline our procedure. 

For each side of WLM, we used a combination of the B$+$C$+$D array and
the D array mosaic rotation curves to fit a mass model. The D array
mosaic rotation curves were generated the same way as the B$+$C$+$D
array rotation curves (see Section~\ref{sec:rotcur}). The D array
mosaic rotation curve extends to larger radii than the B$+$C$+$D array
rotation curve, which is crucial for constraining the halo model. The
B$+$C$+$D array rotation curve is less affected by beam smearing in
the inner regions than the D array mosaic rotation curve, so it is
better at constraining the disk model. We spliced the B$+$C$+$D
rotation curve to the D array mosaic curve at $9\arcmin$ for the
approaching side and at $5\arcmin$ for the receding side. The
B$+$C$+$D array and the D array mosaic rotation curves agree in the
overlap region.

 We used a simple model consisting of a gas disk, a
stellar disk, and a halo to determine the dynamical mass of WLM. For
the gas disk, we found the neutral hydrogen contribution to the total
rotation curve from the HI surface density profile, whose inclination
and position angle were fixed to the appropriate values for the
B$+$C$+$D array data, using the method of \citet{begeman87} and
\citet{casertano83}. The square of the rotation velocity of this
component (which is proportional to the HI mass distribution) was then
multiplied by 1.43 to account for the presence of helium and other
gas. The contribution of the optical component to the total rotation
curve was found from a V-band surface brightness profile
\citep{hunter06} using the same method. We converted the luminosity
distribution of the galaxy into a mass distribution (which is
proportional to the rotation velocity squared) by multiplying the
luminosity distribution by a mass-to-light ratio. The mass-to-light
ratio is a free parameter in our fits. It was restricted to physically
meaningful values, i.e., the mass-to-light ratio had to be greater
than zero. We also assumed that the mass-to-light ratio did not change
as a function of radius in the galaxy. For the halo, we used two
different models: an isothermal sphere and a NFW halo \citep{nfw}. The
first model is generally a good fit to the data and the second model
is suggested by cosmological simulations. The circular velocity for
the isothermal sphere with density distribution
\begin{equation} 
\rho(r) = \rho_0 \left[ 1 + (r/a)^2 \right]^{-1}
\end{equation}
is given by
\begin{equation}
v_{c,iso}(r)^2 = 4 \pi G \rho_0 a^2 [ 1 - (a/r) \arctan (r/a)],
\end{equation}
where $\rho$ is the density, $r$ is the radius, $\rho_0$ is the
central halo density, and $a$ is the core radius
\citep[e.g.][]{mcgaugh1998}. For the NFW halo, the circular velocity
is given by
\begin{equation}
v_{c,nfw}^2 =  \frac{v_{200}^2}{x} \frac{\ln(1 + c x) - c x / (1 + c x)}{\ln(1 + c) - c / (1 + c)}
\end{equation}
where $r_{200}$ is the radius where the overdensity is 200 times that
of the background density, $v_{200}$ is the velocity at $r_{200}$, $x$
is $r/r_{200}$, and c is the concentration parameter. The parameters
$r_{200}$ and $v_{200}$ are related by the equation
\begin{equation}
v_{200} = \left( \frac{G M} { r_{200} } \right)^{1/2} = \left( \frac{r_{200}}{h^{-1} \ \rm{kpc}}\right) \left[ \frac{\Omega_0}{\Omega(z_0)} \right]^{1/2} (1 + z_0)^{3/2} \ \rm{km \ s^{-1}}
\end{equation}
where $\Omega_0$ is the current matter density and $\Omega(z_0)$ is
the matter density at redshift $z_0$. The redshift of WLM was taken to
be 0. The concentration parameter is just $r_{200}$ divided by the
scale radius of the halo so it was restricted to be greater than or
equal to one ($r_{200} \ge r_s$).

We fit three different types of mass models to the data: a maximum
disk model, a maximum halo model, and a best fit mass model. The
maximum disk model fits the data assuming a maximal contribution from
the disk to the rotation curve. Note that this mass model still
includes a halo component to the fit since a disk rotation curve
cannot replicate the flat portion of the observed rotation curve. The
maximum halo model fits the data assuming that there is no
contribution to the rotation curve from the stellar or gas disk. These
two fits, while clearly not physical, provide important boundary
conditions for our mass model of WLM. The best fit mass model is a
model of the galaxy which includes contributions from the halo, the
stellar disk, and the gas disk. For this fit, the parameters for each
component of the mass model are chosen to minimize the difference
between the observed and the model rotation curves using the $\chi^2$
parameter as our minimization criterion. We used the downhill simplex
(amoeba) method of Nelder and Mead \citep{nrfortran} to find the mass
model parameters which minimized the $\chi^2$ value. While this might
not be the best algorithm for minimizing the $\chi^2$ parameter, it
does have the advantage of being easy to implement and relatively fast
given that we only had to fit three parameters. To ensure that the
algorithm found a truly global minimum, we ran it with a grid of
different initial guesses for the parameters and chose the solution
that had the lowest chi-squared value. See
Tables~\ref{tab:iso_massmodel_summary} and
\ref{tab:nfw_massmodel_summary} for a summary of the mass model fits
and Figure~\ref{fig:mass_model} for plots of the best fit mass models.

We were not able to generate an acceptable mass model fit to the
entire rotation curve for the approaching side. The rotation curve for
the approaching side (see the plots on the left hand side of
Figure~\ref{fig:mass_model}) increases rapidly at radii greater than
$8\arcmin$, deviating from the behavior of a well-behaved rotation
curve (cf. the rotation curve for the receding side of the galaxy in
show in the plots on the right hand side of
Figure~\ref{fig:mass_model}). The warp can also be seen in
Figure~\ref{fig:wlm_comb_mom1}. The isovelocity contours form a
U-shape on the northern (approaching) side of the galaxy, while the
contours on the southern (receding) side remain more or less
straight. Unfortunately, there is not enough information in either the
B$+$C$+$D rotation curve or the D array mosaic rotation curve to
constrain the change in inclination and/or position angle. However,
the unusual rotation curve generated for the approaching side obtained
by fixing both the inclination and position angle to a single set of
values points to unconstrained variations in the inclination and
position angle. Since the approaching side of the galaxy is affected
by a warp at radii greater than $8\arcmin$, we excluded these radii
from our fit. Note that the results in
Tables~\ref{tab:iso_massmodel_summary} and
\ref{tab:nfw_massmodel_summary} for the approaching side only go out
to the last point we fitted.

Due to the apparent warp in the approaching side of WLM, we used the
isothermal sphere best fit mass model to the receding side of the
galaxy to determine an estimate for the dynamical mass of the galaxy
($2.16 \times 10^9 \ \rm{M_\odot}$) and the mass model parameters for
the entire galaxy. The maximum halo model (no disk) for the receding
side yields a relatively good fit to the data for the isothermal
sphere case. The maximum disk model, however, yields poor $\chi_r^2$
values for all fits. WLM is clearly not a galaxy whose dynamics are
dominated by its disk. Note that WLM's estimated total dynamical mass
changes by less than 10\% between the isothermal sphere halo and the
NFW halo best fits for the undisturbed receding side of the galaxy.

We also compared our best fit mass model (isothermal sphere halo plus
gas and stellar disks) parameters for WLM with those derived for a
sample of five dwarf irregular galaxies by \citet{cote2000}. We find
that the core radius and central density for the receding side of the
rotation curve as well as the derived galaxy parameters (stellar $M/L$
ratio, total dynamical mass, and ratio of mass in dark matter to the
mass in luminous matter) are within the range of values established by
these authors. The halo parameters for the approaching side of the
galaxy fall outside the \citeauthor{cote2000} distribution, but these
parameters are not well constrained for the the approaching side
because of the warp. WLM, when compared with the galaxies in the
\citeauthor{cote2000} sample, is on the extremely dark
matter-dominated end of the dwarf galaxy spectrum.

\clearpage


\begin{deluxetable}{llll}
\tablewidth{0pt}
\tablecaption{Summary of Observations \label{tab:obs_summary}}
\tablehead{ \colhead{Array Configuration} &
\colhead{B} & 
\colhead{C} & 
\colhead{D} }
\startdata
Date				& 2002 July 9 and 12	& 2000 March 31  &2001  October 19	\\
Number of Pointings		& 1			& 1		 & 9			\\
Square Field of View (\arcmin)	& 30			& 30		 & 60			\\
Channel Width (\kms)		& 2.57 		        & 2.57 	         & 2.57 		\\
Flux Calibrator			& 0137+331		& 0137+331	 & 0137+331		\\
Phase Calibrator		& 2357-114		& 2321-163	 & 2321-163		\\
Total Integration Time (hours)	& 8.5			& 5.1		 & $\sim$2 	        \\
\enddata
\end{deluxetable}

\begin{deluxetable}{lcccc}
\tablewidth{0pt}
\tablecaption{Summary of Final Data Cubes \label{tab:data_summary}}
\tablehead{ 
\colhead{}          &	            & \colhead{Beam Size} & \multicolumn{2}{c}{1$\sigma$ Noise per Channel} \\
\colhead{Cube Name} &	Weighting   &\colhead{(\arcsec)} & \colhead{(mJy beam$^{-1}$)} & \colhead{(K)} } 
\tablecolumns{5}
\startdata
D array mosaic      & Natural & 83.84 $\times$ 59.11	& 4.7	& 0.58  \\
D array mosaic      & Uniform & 72.77 $\times$ 48.88	& 5.1	& 0.87 \\
B$+$C$+$D array cube& Robust=0& 16.17 $\times$ 10.59	& 1.4	& 5.0 \\
\enddata
\end{deluxetable}

\begin{deluxetable}{llcccc}
\tablewidth{0pt}
\tablecaption{Tilted Ring Rotation Curve Fit Results \label{rotcur_results}}
\tablehead{
\colhead{} &
\colhead{} &
\colhead{$<$P.A.$>$} &
\colhead{$\sigma_{P.A.}$} &
\colhead{$<$Incl$>$} &
\colhead{$\sigma_{Incl}$} \\
\colhead{Data Set} &
\colhead{Fit} &
\colhead{($^\circ$)} &
\colhead{($^\circ$)} &
\colhead{($^\circ$)} &
\colhead{($^\circ$)} } 
\tablecolumns{6}
\startdata
D array mosaic &  Entire Galaxy	      &   179.3		&   2.2
&  69.8		& 3.7  \\
D array mosaic &  Approaching	      &   174.6		&  2.3		& 62.8		& 6.0   \\
D array mosaic &  Receding	      &   179.5		&  2.0		& 72.2		& 3.2   \\
B$+$C$+$D array &  Approaching        &  175.5		&  3.8		& 74.2		& 3.1   \\
B$+$C$+$D array &  Receding	      &  172.3		&  4.1		& 82.4		& 5.8  \\
\enddata
\end{deluxetable}

\begin{deluxetable}{llllllll}
\tablewidth{0pt}
\tablecaption{Summary of Mass Model Fits with Isothermal Sphere Halos \label{tab:iso_massmodel_summary}}
\tablehead{
\colhead{} &
\colhead{$a$ } & 
\colhead{$\rho_0$} &
\colhead{$M/L$} &
\colhead{} &
\colhead{$M_{gas}$} &
\colhead{$M_{stars}$} &
\colhead{$M_{halo}$} \\
\colhead{Fit Type} &
\colhead{(\arcsec)} &
\colhead{(\Msun pc$^{-3}$)} &
\colhead{(\Msun$/ \rm{L}_\odot$)} &
\colhead{ $\chi^2_r$} &
\colhead{(\Msun)} &
\colhead{(\Msun)} &
\colhead{(\Msun)}}
\tablecolumns{8}
\startdata
\cutinhead{Approaching Side} 
Best Fit  &  71.5	& 0.157   & 0			    & 2.00     & $6.46 \times 10^7$	& 0			& $3.82 \times 10^8$ \\
Max. Halo &  83.3	& 0.150   & \nodata		    & 2.57     &  0			& 0			& $4.75 \times 10^8$ \\
Max. Disk &  0.00985	& 174000  & 4.8			    & 9.90     & $6.46 \times 10^7$     & $2.07 \times 10^8$    & $1.02 \times 10^7$ \\
\cutinhead{Receding Side} 
Best Fit  & 337	        & 0.0265  &  1.1		    & 0.98   & $9.91 \times 10^7$	& $4.73 \times 10^7$    & $2.01 \times 10^9$ \\
Max. Halo & 237		& 0.0483  &  \nodata		    & 0.84   &  0			& 0			& $2.15 \times 10^9$ \\
Max. Disk & 6870	& 0.00737 & 3.6			    & 8.88   & $9.91 \times 10^7$	& $1.55 \times 10^8$    & $2.64 \times 10^9$ \\
\enddata
\end{deluxetable}
\begin{deluxetable}{llllllll}
\tablewidth{0pt}
\tablecaption{Summary of Mass Model Fits with NFW Halos \label{tab:nfw_massmodel_summary}}
\tablehead{
\colhead{} &
\colhead{$v_{200}$ } &
\colhead{} &
\colhead{$M/L$} &
\colhead{} &
\colhead{$M_{gas}$} &
\colhead{$M_{stars}$} &
\colhead{$M_{halo}$} \\
\colhead{Fit Type} &
\colhead{(\kms)} &
\colhead{$c$ } &
\colhead{(\Msun$/ \rm{L}_\odot$)} &
\colhead{ $\chi^2_r$}  &
\colhead{(\Msun)} &
\colhead{(\Msun)} &
\colhead{(\Msun)} }
\tablecolumns{8}
\startdata
\cutinhead{Approaching Side} 
Best Fit  & 24.6  & 6.60    &  0.965	& 2.34    & $6.46 \times 10^7$  & $4.17 \times 10^7$ & $3.24 \times 10^8$  \\
Max. Halo & 32.1  & 6.81    &  \nodata  & 5.42    & 0			& 0		     & $5.07 \times 10^8$  \\
Max. Disk & 3.33  & 2.28    &  4.8	& 10.42   & $6.46 \times 10^7$  & $2.08 \times 10^8$ & $5.82 \times 10^6$  \\
\cutinhead{Receding Side} 
Best Fit  & 149.6	& 1.0	& 0		& 7.73    & $9.91 \times 10^7$  & 0		     & $2.23 \times 10^9$  \\
Max. Halo & 165.4	& 1.0   & \nodata	& 8.55    & 0			& 0		     & $2.50 \times 10^9$  \\
Max. Disk & 87.3	& 1.0   & 3.6		& 63.94   & $9.91 \times 10^7$  & $1.56 \times 10^8$ & $1.26 \times 10^9$  \\
\enddata
\end{deluxetable}


\clearpage

\begin{figure}
\centering
\caption{Emission in each channel for the naturally weighted D array
mosaic of WLM. Every other channel is shown. The contours are 1, 2, 5,
10, 20, and 30 times the 3$\sigma$ noise level (14.1 mJy~beam$^{-1}$;
0.56K). The beam is given in the lower left hand corner of the first
panel. The channel maps for the uniformly weighted D array mosaic of
WLM are very similar. }
\label{fig:chanmap_wlmdnat}
\end{figure}

\begin{figure}
\centering
\caption{Integrated intensity map of WLM from the naturally weighted D
array mosaic. The contour levels are 1, 10, 20, 30, 50, 100, 200, and
300 times the 3$\sigma$ noise level (8.1$\times 10^{18}$
cm$^{-2}$). The greyscale image is provided to help distinguish the
contours. The emission at the edge of the plot is result of the
mosaicking process (we have better sensitivity at the center of the
image due to the overlap of pointings than at the edges) and is not
real emission. \citet{barnes04} find no companion HI clouds near WLM
with a detection limit of $1.9 \times 10^5 \rm{M}_\odot$ for a object
with a velocity dispersion of 10 \kms. The beam is given in the lower
left hand corner of the image.}
\label{fig:wlmd_nat_mom0}
\end{figure}

\begin{figure}
\centering
\caption{Emission in each channel for the B$+$C$+$D array cube. Every
other channel is shown. The baryocentric velocity of the channel is in
the upper right of each panel. The contours are for 3, 10, and
20$\sigma$ emission. The 1$\sigma$ noise is 1.4 mJy~beam$^{-1}$, which
is equal to 5.0K for our beam. The beam is shown in the last panel.}
\label{fig:chanmap_wlmbcd}
\end{figure}

\begin{figure}
\figurenum{3}
\centering
\caption{{\em Continued} }
\end{figure}

\begin{figure}
\centering
\caption{Integrated intensity (moment zero) map of WLM from the
B$+$C$+$D array cube. The contour levels are 3, 20, 45, 90, and 120
times the 1$\sigma$ noise level (2.9$\times 10^{19} \
\rm{cm}^{-2}$). The greyscale image is provided to help distinguish
the contours. The beam is shown in the lower left hand corner of the
image.}
\label{fig:wlm_comb_mom0}
\end{figure}

\begin{figure}
\centering
\caption{Mean velocity map (moment one) of WLM from the B$+$C$+$D
array cube. The contour levels are -150, -140, -130, -120, -110, -100
and -90 \kms. The greyscale image is provided to help distinguish the
contours. The beam is shown in the lower left hand corner of the
image.}
\label{fig:wlm_comb_mom1}
\end{figure}

\begin{figure}
\centering
\caption{Velocity dispersion (moment 2) map of WLM from the
B$+$C$+$D array cube. The contour levels are two and three times the
width of one channel (2.57 \kms). The greyscale image is provided to
help distinguish contours. The beam is shown in the lower left hand
corner of the image.}
\label{fig:wlm_comb_mom2}
\end{figure}

\begin{figure}
\centering
\caption{Integrated HI spectrum of WLM. The errors in this spectrum
  are on the order of the width of the plotted line.}
\label{fig:wlm_hi_spectrum}
\end{figure}

\begin{figure}
\centering
\caption{Surface density profile for WLM (black dots and black
triangles connected with a solid black line) compared to the surface
density profiles of other dwarf galaxies (colored lines). The black
triangles show the surface density profile for the B$+$C$+$D array
cube, while the black dots show the surface density profile for the
naturally weighted D array mosaic cube. $R_D$ is the V band disk scale
length from \citet{hunter04}. For WLM, $R_D = 2.0 \arcmin$. The
references for the surface density profiles of the other galaxies are:
NGC 2366, \citet{hunter_n2366}; NGC 4449, \citet{hunter_n4449}; IC
1613, E.M. Wilcots (2001, private communication); DDO 105 and DDO 168,
\citet{Broeils92}; DDO 154, \citet{carignan_ddo154}; DDO 50,
\citet{puche_hoII}; DDO 88, \citet{simpson05a}; and DDO 43,
\citet{simpson05b}.}
\label{fig:wlm_combined_sd_plot}
\end{figure}

\begin{figure}
\centering
\caption{A combined color image of the UBV optical data with the HI
contours from Figure~\ref{fig:wlm_comb_mom0} overlaid.}
\label{fig:wlm_color_contour}
\end{figure}

\begin{figure}
\centering
\caption{HI contours overlaid on the \ha image of WLM. The contour
levels are 1, 30, 50, 70, 100, 130, 160 times the 1$\sigma$ noise
level (2.9 $\times 10^{19} \rm{cm}^{-2}$). The C1 and C2 \ha complexes
from \citet{hm95} are indicated by arrows. }
\label{fig:wlm_ha_mom0}
\end{figure}

\begin{figure}
\centering
\caption{The location of the \citet{dolphin00} HST WFPC2 fields
overlaid on a DSS image of WLM. The red squares indicate the position
of the \ha regions identified by \citet{hm95}. The C1 and C2 complexes
are directly north and south of the easternmost WFPC2 field. This
image was created using Aladin \citep{aladin}.}
\label{fig:wlm_wfpc2}
\end{figure}

\begin{figure}
\centering
\caption{HI spectra  for each  pixel in the  C2 \ha complex.  Note the
  double peaked profiles near the center of the region. The separation
  between  the two  peaks  is  approximately 9  \kms,  which gives  an
  expansion velocity for the bubble of 4.5 \kms.}
\label{fig:wlm_plcub_hole}
\end{figure}

\begin{figure}
\centering
\caption{Position velocity diagram for a cut along the major axis of
the neutral hydrogen distribution of WLM ($\rm{PA} = 179.6^\circ$,
which is the position angle derived for the D array mosaic in
Section~\ref{sec:rotcur}). Positive offsets are along the receding
axis and negative offsets are along the approaching axis.}
\label{fig:pv_wlm_comb}
\end{figure}

\begin{figure}
\centering
\caption{Summary of tilted ring model rotation curve fits. The top
panel shows the rotation velocity as a function of radius, the middle
panel shows the inclination as a function of radius, and the bottom
panel shows the position angle as a function of radius. The two fits
are shown in different colors: blue lines are for the fit to the
approaching side of WLM using the B$+$C$+$D array cube and red lines
are for the fit to the receding side of WLM using the B$+$C$+$D array
cube. Error bars for each curve are shown in the corresponding
color. The error bars for some points are quite large and extend
vertically all the way across the graph. The thin solid lines running
across the inclination and position angle plots are the weighted
average of the inclinations and position angles for each fit.}
\label{fig:wlm_rotcur_vripa_all}
\end{figure}

\begin{figure}
\centering
\caption{Best fit mass models for WLM. The left hand plots are for the
approaching side of the galaxy and the right hand plots are for the
receding side of the galaxy. The top row shows the fit for an
isothermal sphere halo and the bottom row shows the fit for an NFW
halo. The rotation curve is a combination of the B$+$C$+$D array and
the D array mosaic rotation curves.  The first two points were
excluded from the fits to either side because they are the points most
affected by inclination and beam smearing effects. Points at radii
greater than $8\arcmin$ were excluded from the fit to the approaching
side because beyond this radius the galaxy is warped and we are unable
to constrain its inclination. The dashed line is the contribution to
rotation curve from the best fit halo, the dotted line is the HI
contribution, the dash-dot curve is the total disk contribution, and
the solid line is the total rotation curve (disk plus halo). Note that
because the derived M/L ratio of the disk is zero for some fits, the
HI contribution and the total disk contribution lines overlap and
there is no plotted contribution from the stellar distribution.}
\label{fig:mass_model}
\end{figure}

\end{document}